\documentclass[aps,pre,twocolumn,amsmath,amssymb,showpacs,amsfonts]{revtex4}
\usepackage{epsfig}
\usepackage{graphicx}
\usepackage{dcolumn}
\usepackage{bm}

\begin{document}

\title{Universality of attractors at weak dissipation and particles distribution in turbulence}

\author{Itzhak Fouxon}
\affiliation{Raymond and Beverly Sackler School of Physics and Astronomy,
Tel-Aviv University, Tel-Aviv 69978, Israel}
\begin{abstract}

We study stationary solutions to the continuity equation for weakly compressible flows. These describe non-equilibrium steady states
of weakly dissipative dynamical systems.
Compressibility is a singular perturbation that changes the steady state density from a constant "microcanonical" distribution into a singular multifractal measure supported on the "strange attractor".
We introduce a representation of the latter and show that the space-averaged properties are described universally
by a log-normal distribution determined by a single structure function.
The spectrum of fractal dimensions is derived. Application to the problem of
distribution of particles in turbulence gives testable predictions for real turbulence and stresses the role of pressure
fluctuations.

\end{abstract}
 \maketitle

Phase space density, that describes averages of dynamical variables, is one of the most fundamental objects
of the classical equilibrium statistical mechanics \cite{Ma,Dorfman}. It obeys the continuity or Liouville equation which time-independent solution describes a steady state. For mixing systems the solution is the microcanonical uniform distribution. A constant solution exists due to the incompressibility of the velocity field in the phase space. Here we study a perturbation of the above scheme in the general setting of
weakly dissipative dynamical systems \cite{Dorfman,Ruelle1}. A dynamical system is represented by a point in a $d-$dimensional space which motion is governed by the local value of a prescribed smooth velocity field. Weak dissipation corresponds to small compressibility of the velocity.
We allow for a generic, not necessarily Hamiltonian, incompressible component and the results apply both to a time-independent deterministic flow and to a time-dependent flow which statistics defined by
spatial averaging is stationary.

Compressible perturbation is singular. It transforms the stationary density from a constant to a singular density, traditionally called the Sinai-Ruelle-Bowen (SRB) measure \cite{Ruelle1,Dorfman,Sinai,Bowen}, supported on a multi-fractal. This needs a special approach which is developed here. As in the usual near-equilibrium physics, the properties of the SRB measure at weak compressibility are expressed via the fluctuations in the unperturbed microcanonical ensemble.


The states described by the SRB measure have a finite rate of the Gibbs entropy production.
The rate finiteness does not contradict that in the steady state the entropy should be constant, as the latter is not finite by the measure singularity. Thus the SRB states are a model of non-equilibrium steady
states where the system exchanges entropy with the environment constantly.


An example of a dissipative dynamical system that played a crucial role in the development of understanding of chaos and non-equilibrium physics is
the Lorenz oscillator \cite{Lorenz}. It demonstrated what is now known as a general property of the dissipative dynamics. At large times
the trajectories asymptote a zero-volume, multi-fractal set in space - the "strange attractor". Thus a small perturbation transforming
incompressible velocity into a compressible one leads to a non-small change in the steady state
density: it becomes infinite on its multi-fractal support. Such a dramatic change is due to the infinite time of evolution implied by the consideration of a steady state. While locally compressibility leads to a small disbalance of trajectories going in and out of
phase space regions, the disbalance effect accumulated over a long time is big. The analysis should therefore disentangle the effects of long time of evolution and of small local disbalances that can be tackled as a small perturbation.

For a mixing incompressible velocity the evolution of a small volume of the phase space makes it dense in space. Its coarse-graining over an
arbitrarily small scale covers all the available phase space, assumed finite here. When a small compressible component is added to the velocity, the
coarse-graining of the evolved volume over an arbitrarily small scale does not cover the whole space any longer.
However the coarse-graining over a small but finite scale, that tends to zero analytically with compressibility, already covers the whole volume. Thus at small compressibility the attractor is "almost" dense in space. The gaps in its structure have volume equal to the whole volume of the phase space but a minute coarse-graining is enough to fill the space. It should be stressed the attractor does have a non-trivial structure.

A practically important example of weakly dissipative dynamics is heavy particles in a turbulent flow. Here physical and phase spaces coincide and
one can actually {\it see} the attractor. The latter, like the velocity, evolves in time keeping its space-averaged properties constant.
At small compressibility, universality  allows to describe the particles distribution in real turbulence without knowing the details of the statistics of turbulence itself.

We pass to the analysis of the continuity equation
\begin{eqnarray}&&
\partial_t n+\nabla\cdot(n\bm v)=0,\ \ \label{continuity}
\end{eqnarray}
for the phase-space density $n$ of a general dynamical system which trajectories ${\bf q}(t, {\bf x})$ are governed
by a smooth $d-$dimensional compressible velocity field ${\bf v}$ in a finite domain (which volume we set to unity) according to
\begin{eqnarray}&&
\partial_t {\bf q}(t, {\bf x})={\bf v}[t, {\bf
q}(t, {\bf x})],\ \ {\bf q}(0, {\bf x})={\bf x}. \label{basic1}
\label{main}
\end{eqnarray}
We study both the case of a time-independent deterministic flow $\bm v(t, \bm x)=\bm v(\bm x)$ and the case of a time-dependent $\bm v(t, \bm x)$ stationary with respect to the statistics defined by the spatial averaging. The steady state solution $n_{SRB}$ to Eq.~(\ref{continuity}) is time-independent in the former and time-dependent and stationary in the latter cases. We study
the space averaged characteristics of $n_{SRB}$, that are time-independent in both cases. The analysis
is written for $\bm v=\bm v(\bm x)$ while generalizations to $\bm v=\bm v(t, \bm x)$ case are either straightforward or follow \cite{FF,IF} literally. We assume $\bm v$ is
weakly compressible and its decomposition into transversal $\bm u$ and longitudinal $\bm u'$ components
obeys
\begin{eqnarray}&&
\!\!\!\!\!\!\!\!\!\!{\bf v}\!=\!{\bf u}\!+\!{\bf u}',\ \ \!\nabla\cdot{\bf v}\!=\!
\nabla\cdot{\bf u}'\!\equiv \! w, \ \ \!u'\ll u, \ \ w\ll |\nabla \bm u|.\label{weak}
\end{eqnarray}
The above setting arises naturally for inertial particles in an incompressible turbulent
flow ${\bf u}(t, {\bf x})$ where particle's drag by the flow is describable by the linear
Stokes friction (see \cite{MaxeyRiley} for details). The Newton law reads
\begin{eqnarray}&&
\dot {\bm q}={\bm v},\ \ \tau\dot {\bm v}+\bm v=\bm u(t,\bm q[t]).
\end{eqnarray}
At $\tau\to 0$ the particle follows the flow, $\bm v_0=\bm u(t,\bm q[t])$. For leading order correction
in $\tau$ put $\bm v=\bm v_0$ in $\tau\dot {\bm v}$, 
\begin{eqnarray}&&
\bm v(t)\approx \bm u-\tau[\partial_t\bm u+(\bm u\cdot\nabla)\bm u], \label{StokesAppr}
\end{eqnarray}
where $\dot {\bm v_0}=\partial_t\bm u+(\bm u\cdot\nabla)\bm u$ is the substantial derivative along
the trajectory of the fluid particle and the RHS is evaluated at $\bm x=\bm q(t)$. In this order the
particle velocity is determined by its location in space. One may introduce the velocity field
$\bm v(t, \bm x)\equiv \bm u-\tau[\partial_t\bm u+(\bm u\cdot\nabla)\bm u]$ that determines particles
trajectories in space as in Eq.~(\ref{main}).
The small correction $\bm v-\bm u$ gives $\bm v$ a
finite compressibility
\begin{eqnarray}&&
\omega=-\tau \nabla\cdot [(\bm u\cdot\nabla)\bm u]=\tau\nabla^2 p\neq 0.
\end{eqnarray}
Above we took divergence of the Navier-Stokes equations $\partial_t\bm u+(\bm u\cdot\nabla)\bm u=-\nabla p+\nu\nabla^2 \bm u+\bm f$ where $p$ is the pressure, $\nu$ is the viscosity and the force $\bm f$ is assumed to be either incompressible or concentrated on the boundaries.
One can use only the potential part of the correction, $\bm v= \bm u+\tau\nabla p$, as it becomes clear from the formulas below.
Formation of a singular density is characteristic for particles and it holds already for a centrifuge: particles shift a bit after each rotation eventually reaching the boundary to form a singular density. Incompressible turbulence can be thought of as a collection of vortices-centrifuges where particles tend to accumulate on the boundaries between the vortices, cf. \cite{MaxeyRiley,FFS1,Olla}.

Unless $\bm v$ is degenerate $n_{SRB}$ is non-smooth. The degeneracy was quantified in
\cite{IF}, based on a Green-Kubo type formula for the space-averaged
sums of backward- $\lambda_i^-$ and forward-in-time $\lambda_i^+$ Lyapunov exponents \cite{FF,IF},
\begin{eqnarray}&&
\!\!\!\!\!\!\!\!\!-\sum\lambda_i^{\pm}=\pm \int_0^{\pm\infty}dt\int w(0, \bm x)w[t, \bm q(t, \bm x)] d\bm x.
\end{eqnarray}
At weak compressibility one may substitute $\bm q(t, \bm x)$ above by the trajectories of incompressible flow $\bm X(t, \bm x)$ defined by
$\partial_t {\bm X}(t, \bm x)=\bm u[t, \bm X(t, \bm x)]$ and $\bm X(0, \bm x)=\bm x$.
We define the "microcanonical" correlation function by the usual
\begin{eqnarray}&&
\langle w(0)w(t)\rangle\equiv \int w(0, \bm x)w[t, \bm X(t, \bm x)] d\bm x.
\end{eqnarray}
By incompressibility $\langle w(0)w(t)\rangle=\langle w(0)w(-t)\rangle$ so that
\begin{eqnarray}&&
\!\!\!\!\!\!\!\!\!\sum \lambda_i^-=\sum \lambda_i^+=-\int_{-\infty}^\infty \langle w(0)w(t)\rangle dt/2=-E(0)/2,\nonumber
\end{eqnarray}
where $E(0)$ is the spectrum of $\omega[t, \bm X(t, \bm r)]$ at zero frequency, $E(0)\geq 0$.
Assuming the sums of Lyapunov exponents are equal for all $\bm r$ except for a set of zero volume (e.g. this holds for systems obeying the
SRB theorem \cite{Dorfman,Ruelle1,Sinai,Bowen}), and remembering the sums
determine the rates of growth of density \cite{IF}, we have for the solutions to Eq.~(\ref{continuity}),
\begin{eqnarray}&&
\!\!\!\!\!\!\!\!\!\lim_{t\to\infty}\frac{1}{t}\ln\frac{n[t, \bm q(t, \bm r)]}{n(0, \bm r)}\!=\!-\!\lim_{t\to\infty}\frac{1}{t}\ln\frac{n(t, \bm r)}{n[0, \bm q(0| t, \bm r)]}=\frac{E(0)}{2},\nonumber
\end{eqnarray}
where $\bm q(0| t, \bm r)$ is the trajectory passing via $\bm r$ at time $t$. For smooth $n(t=0)$, the above implies that,
in a non-degenerate case of $E(0)>0$, at large times, $n(t)$ tends to zero at all $\bm r$ except for a set of zero volume, while
becoming infinite on $\bm q(t, \bm r)$. This rules out smooth $n_{SRB}$ and shows a strange attractor is formed for
Eqs.~(\ref{basic1})-(\ref{weak}) generally. It is
immediate to infer the Kaplan-Yorke codimension $\Delta$ of this attractor. At weak compressibility we have $\sum \lambda_i<0$ but $\sum_{i=1}^{d-1}
\lambda_i\approx -\lambda_d^0>0$, where we assume that the Lyapunov exponents $\lambda^0_i$ of $\bm u$ do not vanish identically (this implies
$\lambda_d^0<0$ by $\sum \lambda^0_i=0$). Thus
\begin{eqnarray}&&
\Delta=\sum \lambda_i/\lambda_d\approx E(0)/(2|\lambda_d^0|).
\end{eqnarray}
Applying the above to particles in turbulence, we find
\begin{eqnarray}&&
\!\!\!\!\!\!\!\!\sum \lambda_i=(-\tau^2/2)\int_{-\infty}^\infty \langle \phi(0)\phi(t)\rangle dt=-\tau^2 E_{\phi}(0)/2<0,\nonumber
\end{eqnarray}
where $E_{\phi}(0)>0$ is the spectrum of 
$\phi\equiv \nabla^2 p$ and we used that there is no degeneracy.
Thus inertial particles in turbulence asymptote a fractal set with the Kaplan-Yorke codimension
$\Delta=\tau^2 E_{\phi}(0)/(2|\lambda_3^0|)$ where $\lambda_3^0<0$ is the third Lyapunov exponent of turbulence \cite{FFS1,Collins}.

Phase space density serves to find averages of dynamical variables - functions on the phase space, so
only integrals of $n_{SRB}$ with smooth test functions are observable. Thus "weak solutions" to the continuity equation -
possibly singular $n_{SRB}(\bm r)$ satisfying $\int d\bm r n_{SRB}(\bm r) \bm v(\bm r)\cdot \nabla f(\bm r)=0$
for all smooth $f(\bm r)$ (coinciding with $\nabla \cdot[n_{SRB}\bm v]=0$ if $n_{SRB}$ is smooth) describe physically meaningful states. Our main assumption is that such solutions are obtainable as a long-time limit $T\to\infty$ of evolution of a unit initial density set at $t=-T$:
\begin{eqnarray}&&
\!\!\!\!\!\!\!\!\!\!\!\!\!\!n_{SRB}(\bm r)\!=\!\!\lim_{T\to\infty}e^{\rho(-T, \bm r)},\ \ \rho(t, \bm r)\!\equiv \!\!\int_0^t\! w[\bm q(t', \bm r)]dt'. \label{representation}
\end{eqnarray}
The factor $\rho(t, \bm r)$
describes changes in an infinitesimal volume $V(t)$ located at $t=0$ near $\bm r$, as $\rho(t, \bm r)=\ln[V(t)/V(0)]$
where $t$ can be both positive and negative. 
In $\lim_{T\to\infty} \int d\bm r e^{\rho(-T, \bm r)}(\bm r) \bm v(\bm r)\cdot \nabla f(\bm r)$
first one takes the integral over $\bm r$ and
then the limit $T\to\infty$. Then Eq.~(\ref{representation}) defines a weak solution $n_{SRB}$ if for any smooth $f$,
\begin{eqnarray}&&
\!\!\!\!\!\!\!\!\!\!\!\!\!\!\lim_{T\to\infty} I_f(T)\!=\!0,\ \ I_f(T)\!\equiv\!\!\int d\bm r e^{\rho(-T, \bm r)}\bm v(\bm r)\nabla f(\bm r). \label{assumption}
\end{eqnarray}
Changing variables $\bm r=\bm q(T, \bm x)$, using $v_i[\bm q(T, \bm x)]=v_j(x)\nabla_jq_i(T, \bm x)$, see \cite{FF}, and integrating by parts, one finds
$I_f(T)=-\langle \omega(0) f(T)\rangle$. 
We will assume that correlations decay $\langle \omega(0) f(T)\rangle=\langle \omega(0)\rangle\langle f(T)\rangle=0$ at $T\to\infty$ and
$\omega[\bm q(t, \bm r)]$ has a finite correlation time $\tau_c$. Then Eq.~(\ref{representation}) defines a weak solution to the continuity equation.
This obeys
the SRB theorem \cite{Dorfman} on the equality of time and phase-space averages for dynamical variables $f$,
\begin{eqnarray}&&
\!\!\!\!\!\!\!\!\!\!\lim_{T\to\infty} (1/T)\int_0^T f[\bm q(t, \bm x)]dt=\int f(\bm r) n_{SRB}(\bm r)d\bm r,
\end{eqnarray}
provided the time-average is independent of $\bm x$ for almost every $\rm x$. The proof is based on the identity
$\lim_{T\to\infty} \int f[\bm q(T, \bm x)]d\bm x=\int
f(\bm r) n_{SRB}(\bm r)d\bm r$.
The density defined by Eq.~(\ref{representation}) 
is
an exponent of a sum of a large, asymptotically infinite, number $\sim T/\tau_c$ of uncorrelated random variables. This implies $n_{SRB}$
is singular pointwise. 
However, its correlation $\varphi(\bm r)$ at $r\neq 0$ is finite,
\begin{eqnarray}&&
\!\!\!\!\!\!\!\!\!\!\!\!\!\!\!\!\!\!\!\varphi(\bm r)\!\equiv\! \langle n_{SRB}(0)n_{SRB}(\bm r)\rangle\!=\!\!\lim_{T\to\infty}\!\!\Big\langle e^{\rho(-T, 0)+\rho(-T, \bm r)}
\Big\rangle, \label{pairepresentation}\end{eqnarray}
where the angular brackets designate spatial averaging.
The cumulant expansion theorem \cite{Ma,FFS}, applied to the volume conservation $\langle \exp[\rho(t, \bm r)]\rangle=1$, gives the "sum rule"
$\sum_{n=1}^{\infty} \left\langle \rho^n(t, \bm r)
\right\rangle_c/n!=0$,
where the subscript $c$ stands for the cumulant (giving to lowest order in compressibility
$-\langle \rho\rangle=\langle \rho^2\rangle_c/2$).
Applying the expansion to Eq.~(\ref{pairepresentation}) and
using the sum rule, one finds only the mixed terms, containing
both $\rho(t, \bm r)$ and $\rho(t, 0)$ do not cancel,
\begin{eqnarray}&&
\!\!\!\!\!\!\!\!\!\!\!\!\!\!\ln \varphi(\bm r)\!=\!\lim_{T\to\infty}\sum_{n=2}^{\infty}[\langle[\rho_1\!+\!\rho_2]^n\rangle_c\!-\!\langle \rho_1^n\rangle_c\!-\!\langle \rho_2^n\rangle_c]/n!,\label{general}
\end{eqnarray}
where $\rho_1\equiv \rho(-T, 0)$ and $\rho_2\equiv \rho(-T, \bm r)$.
The above leads to a series representation for the attractor's correlation dimension, cf. \cite{MehligWilkinson}. In the limit of
small compressibility,
\begin{eqnarray}&&
\!\!\!\!\!\!\!\!\!\!\!\!\!\!\varphi(\bm r)\!=\!e^{g(\bm r)},\ \  
\!\!g(\bm r)\!\!\equiv\!\!\! \int_{-\infty}^0\!\!\!\!\! dt_1dt_2\langle
\omega[{\bf X}(t_1, 0)] \omega[{\bf X}(t_2, {\bf r})]\rangle. \label{structure} \end{eqnarray}
Above we put $\bm X(t, \bm r)$ instead of $\bm q(t, \bm r)$ and omitted the subscript c since for incompressible
flow $\langle w[{\bf X}(t, {\bf r})]\rangle=\langle w({\bf r})\rangle$ and $\langle w({\bf r})\rangle=\int  \nabla\cdot\bm v d\bm r =0$.
The condition of applicability of Eqs.~(\ref{structure}) is the negligibility of $n>2$ terms in Eq.~(\ref{general}).
Proceeding for $\langle n(\bm r_1)n(\bm r_2)..n(\bm r_N)\rangle=
\lim_{T\to\infty}\langle \exp[\sum \rho(-T, \bm r_i)]\rangle$
as for $\varphi(r)$,
\begin{eqnarray}&&
\!\!\!\!\!\!\!\!\!\!\!\!\!\!\!\ln \langle n_{SRB}(\bm r_1)n_{SRB}(\bm r_2)..n_{SRB}(\bm r_N)\rangle
=\sum_{i>j}g({\bf r}_{i}-{\bf
r}_j). \label{bueno}\end{eqnarray}
Thus $n_{SRB}$ has log-normal statistics completely determined by a single structure function
$g(\bm r)$. The latter behaves universally at small $r$. It diverges at $r=0$  
because $\bm X(t, 0)$ and $\bm X(t, \bm r)$ do not separate and spend infinite time together. Thus
$\langle n_{SRB}^2\rangle =\infty$ reflecting $n_{SRB}$ is singular and compressibility is a
singular perturbation.
At small but finite $r$ the trajectories separate exponentially with the Lyapunov exponent
$|\lambda_d^0|$ of the time-reversed flow $\bm u$. The trajectories
stay together during a time that diverges logarithmically at $r\to 0$. To single out this divergence fix a scale
$r_*\ll\eta$, where below $\eta$ both $\bm v$ and $\omega$ are smooth, and consider $r$ such that $t_*\equiv
\ln[r_*/r]/|\lambda_d^0|\gg \tau_c$. Separating the interval of integration over $t_1$ in Eq.~(\ref{structure}) into $(-\infty, -t_*)$ and $(-t_*, 0)$,
\begin{eqnarray}&&
\!\!\!\!\!\!\!\!\!\!\!\!\!\!\!g(\bm r)\!=\!\!\!\int_{-\infty}^{-t_*}\!\!\!dt_1\!\!\!\int_{-\infty}^{\infty}\!\!\!\!\!dt_2 \langle
\omega[{\bf X}(t_1, 0)] \omega[{\bf X}(t_2, {\bf r})]\rangle\!+\!t_*E(0). \label{decomposition}
\end{eqnarray}
The first term on the RHS is estimated as $g(r_*)$ while the divergence
at small $r$ is due to the last term:
\begin{eqnarray}&&
\varphi(\bm r)\approx\left(r_*/r\right)^{2\Delta}.
\end{eqnarray}
Thus the correlation codimension is twice the Kaplan-Yorke one.
Note $\eta$ is defined up to a factor of order unity but this is of no effect for the answer
by $\Delta\ll 1$. The statistics is isotropic at small $\bm r$ independently of isotropy of $\bm v$. The origin of this is that the separation vector between two infinitesimally close trajectories grows with an exponent independent of its initial direction.

Applying the above to the inertial particles, the statistics is completely determined by the
structure function
\begin{eqnarray}
S(\bm r)\equiv \int_{-\infty}^0\!\!\!\! dt_1dt_2\langle
\phi[t_1, {\bf X}(t_1, 0)] \phi[t_2, {\bf X}(t_2, {\bf r})]\rangle, \nonumber\end{eqnarray}
involving only the properties of turbulence. For density,
\begin{eqnarray}&&
\langle n_{SRB}(0)n_{SRB}(\bm r)\rangle=\exp[\tau^2 S(\bm r)]. \label{pairanswer1}
\end{eqnarray}
To analyze the domain of validity of Eq.~(\ref{pairanswer1}) consider $n=3$ term in Eq.~(\ref{general}).
It is of the type $\tau^3\int dt_1dt_2dt_3 \langle
\phi[t_1, {\bf X}(t_1, 0)] \phi[t_2, {\bf X}(t_2, 0)] \phi[t_2, {\bf X}(t_2, {\bf r})]\rangle_c$.  Its evaluation within Kolmogorov's theory (KT) \cite{Frisch} shows it is smaller than $\tau^2 S(\bm r)$ by the factor of the Stokes number ${\rm St}\equiv \tau |\lambda_3^0|$, the product of $\tau$ and a typical value of the velocity gradient. This leads to the validity condition ${\rm St}\ll 1$ of Eq.~(\ref{pairanswer1}).
Due to the deviations from KT, higher $n$ terms in Eq.~(\ref{general}) contain additional powers of the Reynolds number ${\rm Re}$ and the true condition is ${\rm St} {\rm Re}^{\alpha}\ll 1$ with $\alpha>0$. One expects $\alpha \ll 1$ so in practical situations
${\rm St}\ll 1$ ensures Eq.~(\ref{pairanswer1}). The same condition is expected to satisfy the additional demand that the underlying Eq.~(\ref{StokesAppr}) holds.
For particles, $\eta\sim (\nu/|\lambda_3^0|)^{1/2}$ is the Kolmogorov scale of turbulence where $\nu$ the fluid viscosity \cite{Frisch}. Neglecting intermittency, $\tau^2S(\eta)\sim {\rm St}^2$ so that
$g(\eta)\ll 1$ and one can neglect the first term in Eq.~(\ref{decomposition}) for $t_*=\ln(\eta/r)/|\lambda_3^0|$,
\begin{eqnarray}&&
\!\!\!\!\!\!\!\!\!\!\!\!\langle n_{SRB}(0)n_{SRB}(\bm r)\rangle=\left(\eta/r\right)^{2\Delta},\ \ r\ll \eta. \label{smallscales}
\end{eqnarray}
The above was obtained in \cite{FFS1}. There are no significant density fluctuations,
$\langle n_{SRB}(0)n_{SRB}(\bm r)\rangle\approx \langle n_{SRB}\rangle^2
=1$ already at
$r\ll \eta$, where $\Delta\ln \left(\eta/r\right)\ll 1$.
Turbulence effect on the particles
depends on the observer's resolution: at $\Delta\ln \left(\eta/r\right)\gtrsim 1$
segregation holds, while at larger $r$ - mixing. This is how the assumption of mixing can work effectively for particles on a multifractal. Segregation can also bring physical effects regardless the observer's resolution by influencing the rate of collisions
\cite{FFS1}.

At ${\rm St}\ll 1$ density inhomogeneities are absent in the inertial range $\eta\lesssim r\lesssim L$, where $L$ is the pumping scale of turbulence \cite{Frisch}. Then Eq.~(\ref{smallscales}) is a complete description. In contrast, at ${\rm St}\sim 1$ the inertial range inhomogeneities are important \cite{Stefano} and our  Eq.~(\ref{pairanswer1}), extended to hold asymptotically at ${\rm St}\sim 1$, gives a unique access to the inhomogeneities.
In KT $S(\bm r)$ depends only on $r$ and the mean energy dissipation $\epsilon$
so $\ln \varphi_{KT}(r)\approx C(\tau/\tau^{KT}_r)^2$, where $C$ is dimensionless constant
and $\tau^{KT}_r\equiv \epsilon^{-1/3}r^{2/3}$ is the Kolmogorov time-scale of turbulent fluctuations at scale $r$. The
KT prediction applies rigourously to the Kraichnan model of ${\bf v}$ decorrelated in time \cite{BFF,Collins,BecCenciniHillerbranddelta}.
However for turbulence the numerical simulations \cite{Stefano}
show $\ln \varphi(r)\propto r^{-10/3}$. This is at moderate ${\rm Re}$ where KT is expected to work.
Noting $S(\bm r)\sim \tau_r^2 \partial_r^4\langle
[p({\bf x}+{\bf r})-p({\bf x})]^2\rangle$, where $\tau_r$ is the relevant time-scale, we suggest the difference has the same origin as the deviations of the pressure scaling from KT \cite{Gotoh}.


The absence of fluctuations beyond $\Delta \ln(\eta/r)\gtrsim 1$ 
is the central property of weakly dissipative dynamical systems. Consider the density $n_l$ coarse-grained over the volume $V_l$ of a ball with radius $l$ in $d$ dimensions,
\begin{eqnarray}&&
\!\!\!\!\!\!\!\!\!n_l({\bf r})=m_l(\bm r)/V_l,\ \ m_l\equiv \int_{|{\bf r}'-{\bf r}|<l} n_{SRB}({\bf r'}) d{\bf
r}',
\end{eqnarray}
The mass $m_l$ is equal to the mass contained in the dynamical image
of the ball at time $-t$ which is an ellipsoid with largest axis growing as $l\exp[|\lambda_d^0|t]$ (fluctuations are negligible by weak compressibility). For $-t=t_*\equiv \ln(R/l)/|\lambda_d^0|$ where
$R\ll \eta$, but $\Delta \ln (\eta/R)\ll 1$, this mass is spread over a region where the density self-averages and
it equals just the volume of that region,
$V_l\exp[\rho(-t_*, \bm r)]$,
 \begin{eqnarray}&&
n_l(\bm r)=\exp[\rho(-t_*, \bm r)].
\end{eqnarray}
This is a fundamental formula of the weakly dissipative regime: density fluctuates at scale $l$ due to mass condensing from
a volume with size smaller than $\eta$ over which the mass is effectively distributed uniformly. Using $-\langle \rho^2(-t)\rangle_c/2=\langle \rho(-t)\rangle=t\sum \lambda_i^-$ and $(\eta/R)^{2\Delta}\approx 1$, one finds the lognormal statistics
$\langle n_l^{\gamma}\rangle=\left(\eta/l\right)^{\Delta\gamma(\gamma-1)}$, cf. Eq.~(\ref{bueno}).
To find the fractal dimensions $D(\alpha)$ \cite{HentshelProccacia,BecGawedzkiHorvai},
\begin{eqnarray}&&
D(\alpha)\!\equiv\!\lim_{l\to 0}\ln \langle m_l^{\alpha-1}n_{SRB}\rangle/[(\alpha-1)\ln l]
\end{eqnarray}
consider
$\langle n_l^{\alpha-1}n_{SRB}\rangle=\lim_{T\to\infty}\!
\langle \exp[\alpha\rho(-t_*, \bm r)\!+\!\int_{-T}^{-t_*}\!\!\omega[t, \bm q(t, \bm r)]dt]\rangle$.
For $t_*\gg \tau_c$, using $\rho(-t_*, \bm r)\approx \rho(-t_*+\tau_c, \bm r)$  one
can make independent averaging $\langle \exp\left[\alpha\rho(-t_*, \bm r)+\int_{-T}^{-t_*}\omega(\bm q(t, \bm r)\right]\rangle=\langle
\exp\left[\alpha\rho(-t, \bm r)\right]\rangle\langle \exp\left[\int_{-T}^{-t}\omega(\bm q(t, \bm r)\right]\rangle=\langle n_l^{\alpha}\rangle$,
where we used volume conservation. 
This gives $D(\alpha)=d-\Delta\alpha$: fractal dimensions are close to $d$ and the attractor is almost space-filling (for $\alpha\gg 1$ and high moments the lognormal approximation is not valid generally).
The general result of \cite{BecGawedzkiHorvai} for $D(\alpha)$ in the Kraichnan model in $d=2$ reduces to our result at small compressibility.

Our results give testable predictions for particles in real turbulence with no need for
assumptions about the turbulence structure. 
The fractal structure, with dimensions defined by a single parameter $\Delta$, exists below $r_{cor}$ and is formed in a characteristic time $\ln(\eta/r_{cor})/|\lambda_3^0|\sim |\lambda_3^0|^2/{\rm St}^2 E_{\phi}(0)$. An everyday application is the distribution of water droplets in warm clouds, important for such
problems as radiation or rain formation in clouds \cite{FFS1}. The analysis can be extended to include gravity and
the case of light particles and to describe correlations between densities of different size ($\tau$) particles. 

The author is grateful to M. Wilkinson, M. Cencini, A. Leshansky, J. Bec, R. Vilela, K. Gawedzki, G. Falkovich and J. Kurchan for discussions.
This work was supported by COST Action MP$0806$.

\end{document}